\documentclass[11pt]{article}
\pdfoutput=1

\usepackage{euscript}
\usepackage{amsfonts}
\usepackage{amsbsy}
\usepackage{epsfig}
\usepackage{amsthm}
\usepackage{amscd}
\usepackage{amstext}
\usepackage{verbatim}
\usepackage{cancel}
\usepackage{capt-of}
\usepackage{empheq}

\usepackage[nosort]{cite}
\usepackage{bm}
\usepackage{authblk}
\usepackage{esint}
\usepackage[T1]{fontenc}
\usepackage{mathdots}
\usepackage[centertableaux]{ytableau}

\usepackage[utf8]{inputenc}
\usepackage{graphicx}
\usepackage{physics}
\usepackage{mathtools}
\usepackage{bbold}

\usepackage{setspace}
\usepackage{colortbl}
\usepackage{xcolor}

\usepackage{amsmath}
\makeatletter
\renewcommand*\env@matrix[1][\arraystretch]{%
  \edef\arraystretch{#1}%
  \hskip -\arraycolsep
  \let\@ifnextchar\new@ifnextchar
  \array{*\c@MaxMatrixCols c}}
\makeatother
\usepackage{amssymb}
\usepackage{mathrsfs}
\usepackage{booktabs}
\usepackage{bbm}
\usepackage{float}

\usepackage{pgfplots}
\pgfplotsset{compat=1.15}
\usepackage{tikz}
\usetikzlibrary{external}
\usetikzlibrary{arrows}

\usepackage{subcaption}
\usepackage{graphicx}
\usepackage{mathtools}
\usepackage[many]{tcolorbox}
\tcbset{shield externalize}
\usepackage{xcolor, colortbl}
\usepackage[a4paper]{geometry}
\usepackage[hidelinks,linktocpage]{hyperref}
\hypersetup{
    colorlinks=true,
    linkcolor=blue,
    citecolor=blue,
    }

\usepackage{mathrsfs}

\def\ben{\begin{equation}}
\def\een{\end{equation}}

\let\a=\alpha    
   \let\k=\kappa
  \let\n=\nu

\let\w=\omega

\let\pa=\partial
\def\be{\begin{equation}}
\def\ee{\end{equation}}
\def\beq{\begin{equation}}
\def\eeq{\end{equation}}
\def\ba{\begin{array}}
\def\ea{\end{array}}

\def\dalemb#1#2{{\vbox{\hrule height .#2pt
       \hbox{\vrule width.#2pt height#1pt \kern#1pt
               \vrule width.#2pt}
       \hrule height.#2pt}}}

\newcommand{\bea}{\begin{eqnarray}}
\newcommand{\eea}{\end{eqnarray}}

\def\vep{{\varepsilon}}

\makeatletter
\newcommand*\bigcdot{\mathpalette\bigcdot@{.5}}
\newcommand*\bigcdot@[2]{\mathbin{\vcenter{\hbox{\scalebox{#2}{$\m@th#1\bullet$}}}}}
\makeatother

\title{Bootstrapping transport in the Drude-Kadanoff-Martin model}
\date{\vspace{-5ex}}
\author{Subham Dutta Chowdhury$^\sharp$, Sean A. Hartnoll$^\flat$, Aditya Hebbar$^\sharp$ and Ruby Khondaker$^{\flat\natural}$}
\affil{
{\it $^\sharp$ The Abdus Salam ICTP,
Strada Costiera 11, 34151, Trieste, Italy} \\
{\it $^\flat$ DAMTP, University of Cambridge, Cambridge CB3 0WA, UK} \\
{\it $^\natural$ Rudolf Peierls CTP, University of Oxford, Oxford OX1 3PU, UK} 
}

\begin{document}

\maketitle

\begin{abstract}

The Drude-Kadanoff-Martin (DKM) model is a simple low energy and long wavelength description of charge transport, parameterised by the current relaxation timescale $\tau$, charge diffusivity $D$ and charge compressibility $\chi$. We obtain sharp constraints on these parameters in terms of the microscopic energy and length scales of any underlying lattice model with local and bounded interactions. Our primary tools are upper bounds on the retarded Green's function for the charge density in such a setting. We first note that the DKM model cannot pertain at microscopic energy scales because it is inconsistent with the exponential suppression of spectral weight at the highest frequencies in a lattice model. Secondly, under the assumption that the low energy dynamics is captured by the model, we obtain a lower bound on the collective mean free path $\ell \equiv \sqrt{\tau D}$. This bound is shown to imply a version of the Mott-Ioffe-Regel bound: systems with $\ell$ much shorter than the lattice length scale cannot have conventional Drude peaks, in the sense of being captured by the DKM model.

\end{abstract}

\section{Introduction: transport and bootstrap bounds}

The low energy transport of conserved charges is a ubiquitous physical process, arising in settings as diverse as solid state physics and astrophysics. The parameters that control transport, for example diffusivities, are determined by decay rates and mean free paths in the underlying microscopic `high energy' physics of the system. If the microscopic description is weakly interacting, then the Boltzmann equation provides a formalism to compute transport coefficients from microscopic scattering, although subtleties abound \cite{ziman}. Beyond weak interactions, connecting the low energy and high energy physics of transport remains very challenging.

A longstanding strategy is to identify fundamental bounds on transport coefficients. These have a similar role to the uncertainty relation bounds in quantum mechanics. That is, they identify core physical mechanisms at work and give intuition about how complicated systems will behave. Within solid state physics, a widely discussed notion is that mean free paths should not be shorter than the lattice spacing. For electronic transport in metals this idea is known as the Mott-Ioffe-Regel (MIR) bound \cite{ioffe1960non, mott1972conduction}, see \cite{RevModPhys.75.1085, Hussey} for reviews. The same mean free path bound has been noted for heat transport in insulators \cite{PhysRev.75.972, slack, PhysRevB.29.2884, PhysRevB.46.6131} and was called the Slack-Kittel bound in \cite{Mousatov:2019vgj}. Metallic systems whose conductivities appear to violate the MIR bound have been called `bad metals' \cite{PhysRevLett.74.3253}. Such systems are important because the violation likely indicates that their conductivity is not related to a microscopic mean free path in the conventional way.

Interest in transport bounds was re-ignited two decades ago by the conjectured KSS lower bound on the viscosity \cite{Kovtun:2004de}. Zaanen suggested that the viscosity bound should be considered together with the widely observed $T$-linear resistivity of unconventional metals \cite{Bruin:2013qlp} to hint at a `Planckian' timescale bound $\tau \gtrsim 1/T$, with $T$ the temperature \cite{Zaanen2004}. This bound relates to profoundly different physics than the mean free path bound discussed above --- instead of the ability to form coherent wavepackets it is concerned with bounds on the timescale of quantum many-body thermalisation. This distinction was first emphasised in \cite{Hartnoll:2014lpa}, with a more precise discussion recently given in \cite{Hartnoll:2021ydi}. Planckian dissipation has natural connections to quantum criticality \cite{subirbook} and has been observed in important strongly correlated systems including low temperature cuprates \cite{Legros2019} and magic-angle graphene \cite{PhysRevLett.124.076801}. See \cite{Hartnoll:2021ydi} for a review.

Given the importance of transport bounds, one would like to prove them. Any proof must be based on general physical principles. One set of results has followed from light-cone causality. Such causality arises either as a consequence of Lorentz invariance or from the Lieb-Robinson velocity of local lattice Hamiltonians \cite{Lieb:1972wy, Chen_2023}. The simplest transport process, diffusion, is acausal unless the diffusivity obeys an upper bound \cite{Baier:2007ix, Hartman:2017hhp, Han:2018hlj, Heller:2022ejw, Heller:2023jtd}. A second fruitful approach has been anchored in the fluctuation-dissipation theorem, which implies that transport coefficients (dissipation) also control aspects of the variance in quantum states (fluctuation). If the fluctuations become too large the system is unable to self-consistently establish local thermal equilibrium. This leads to bounds on transport coefficients \cite{Kovtun:2011np, Kovtun:2014nsa, Delacretaz:2018cfk, Delacretaz:2023pxm}. Other approaches include using analyticity properties of hydrodynamic correlation functions \cite{Grozdanov:2020koi} and uncertainty relations applied to microscopic observables in a thermal state \cite{NUSSINOV20251354755}.

In this paper we will take a different approach, inspired by the bootstrap program for scattering amplitudes, which we now outline. Low energy scattering amplitudes are described by an effective field theory that depends on a small number of parameters. A priori it might appear that these parameters could take any value. However, the low energy amplitudes are part of a full S-matrix that is subject to several key constraints. Firstly, unitarity of the S-matrix implies that the imaginary part of the partial waves $T_l(s)$ have both a lower and upper bound, $0 \leq \text{Im}\, T_l(s) \leq 2$, for Mandelstam $s$ variable in the physical region. Secondly, causality implies that $T_l(s)$ is analytic in $s$ over some domain. Thirdly, the existence of a mass gap and a field-theoretic UV completion implies the additional Froissart-Martin \cite{Froissart:1961ux, Martin:1962rt} upper bound on $\left|\sum_l (2l +1)T_l(s) \right|$ at large $s$.
The scattering amplitude bootstrap combines these constraints with an assumed low energy effective field theory to obtain bounds on the parameters controlling low energy scattering. Perhaps the most well-known example is the proof that the low energy Wilson coefficient of the 4 derivative operator is positive \cite{Adams:2006sv}. Similar arguments had been used to constrain parameters in chiral perturbation theory \cite{Pham:1985cr, Pennington:1994kc, Ananthanarayan:1994hf} and were used in the proof of the a-theorem \cite{Komargodski:2011vj, Luty:2012ww}, among many other applications. See \cite{deRham:2022hpx} for a recent review.

We will adapt the bootstrap logic to transport by considering the retarded Green's function for the charge density operator, $G^R(\omega,k)$. This object will play the role of the partial waves in the previous paragraph. The retarded Green's function has positive imaginary part at real frequencies, $0 \leq \text{Im}\, G^R(\omega,k)$, and is analytic in the upper-half complex $\omega$ plane, see e.g.~\cite{Hartnoll:2009sz}. These properties are clearly analogous to two of the partial wave properties that we have just mentioned. Indeed, they have recently been used to show the positivity of Wilson coefficients in theories with spontaneously broken Lorentz symmetry \cite{Creminelli:2022onn, Creminelli:2024lhd}. Unlike partial waves, however, the retarded Green's function does not in general have an upper bound. Upper bounds on partial waves have been an essential ingredient in the modern non-perturbative S-matrix bootstrap program \cite{Paulos:2016but, Paulos:2016fap, Paulos:2017fhb}, leading to constraints on Wilson coefficients in QCD flux tubes \cite{EliasMiro:2019kyf}, supergravity \cite{Guerrieri:2021ivu, Guerrieri:2022sod} and more. For a comprehensive list see \cite{Kruczenski:2022lot}.

In \S\ref{sec:bounds} we obtain upper bounds on $\text{Im}\, G^R(\omega,k)$, integrated over a region of $(\omega,k)$ space. Within the integration domain the Green's function is determined by a low energy effective theory of charge transport. The upper bounds are obtained by requiring the microscopic `high energy' completion of the theory to be a lattice model with local and bounded interactions. This is the natural setting for much of condensed matter physics but has not been considered in the predominantly particle physics oriented bootstrap studies (other aspects of many-body lattice models have been bootstrapped in \cite{xizhi, Cho:2022lcj, Nancarrow:2022wdr}). The first of our bounds has a similar flavour to the $f$-sum rule that pertains in lattice models \cite{PhysRevB.16.2437, RevModPhys.83.471} --- in essence we trade a stronger frequency resolution, relative to the sum rule that integrates over all frequencies, for an inequality rather than an equality. The second is a version of the operator growth bounds in \cite{PhysRevLett.115.256803, PhysRevX.9.041017}.

In \S\ref{sec:drude} we pick an effective low energy and long wavelength description of charge transport, which we call the Drude-Kadanoff-Martin (DKM) model. The model consists of diffusive charge dynamics supplemented by a single current relaxation timescale \cite{KADANOFF1963419}. Using this model to compute the integrals, our bounds constrain low energy transport parameters in terms of microscopic length and energy scales. In \S\ref{sec:micro} we make the elementary point that the power-law falloff of spectral weight at high frequencies in the DKM model is not compatible with the exponential falloff required by the operator growth bound. The DKM model, therefore, cannot be a microscopic model. In \S\ref{sec:free} we show that the constraints on $\text{Im}\, G^R(\omega,k)$ require the DKM model to obey a version of the MIR bound, connecting also to widely discussed issues around Planckian dissipation and resistivity saturation. While our results are consistent with general expectations, it should be emphasised that the bounds we obtain are numerically precise subject to sharply defined assumptions about the effective low energy description.

The bootstrap logic, that we are following, takes the effective description as given. The bounds exclude certain microscopic theories as possible short distance completions of the effective description, but do not aim to prove that the effective description actually arises from any given microscopic Hamiltonian. To obtain our bounds we make the fairly strong assumption that the DKM model pertains over a range of frequencies and wavelengths, such that the Drude peak in the optical conductivity is fully contained within the low energy description. This is not the case for many interesting materials, whose peaks exhibit multiple timescales \cite{armitage}. In \S\ref{sec:discussion} we discuss extensions of our approach beyond the simple DKM model. As we explain, it will be possible to establish more general bounds by leveraging the analyticity properties of $G^R(\omega,k)$, which have otherwise been under-utilised in the present paper.

\section{Bounds on integrated spectral weight}
\label{sec:bounds}

Let $n_\a$ be the charge operator at the $\a$th site of a translation-invariant $d$-dimensional lattice model. We can define the charge density on the lattice as $n(\vec x_\a)  \equiv \frac{n_\a}{\nu}$,
where $\vec x_\a$ is the position of the $\a$th lattice site and $\nu$ is the volume of the unit cell in the lattice. The charge density is important because it survives as an operator in the effective continuum long-wavelength description. The retarded Green's function for the charge density is then
\be\label{eq:gr}
G^R(\omega,\vec k) \equiv \frac{i}{\nu} \sum_\a \int_0^\infty dt \, e^{i \omega t - i \vec k \cdot \vec x_\a } \tr \left([n_\a(t),n_0(0)]  \frac{e^{- H/T}}{{\mathcal Z}_T} \right) \,,
\ee
where ${\mathcal Z}_T \equiv \tr e^{-H/T}$ is the partition function, with $T$ the temperature and $H$ the Hamiltonian.
While there are two factors of the density in the Green's function (\ref{eq:gr}), one of the inverse factors of $\nu$ is removed by
the continuum integral over space, $\int d^dx = \nu \sum_\a$.

In the methods section \ref{sec:methods} we derive two upper bounds on the spectral weight $\text{Im} \, G^R(\omega,\vec k)$. The spectral weight will be integrated over some sub-region $\Sigma$ of the Brillouin zone and over a range of frequencies from $\omega$ to a cutoff $\Lambda$. These bounded domains are important as we will only assume the form of the Green's function at low energies and long wavelegnths. Firstly, using similar physical input to the standard $f$-sum rule \cite{PhysRevB.16.2437, RevModPhys.83.471} we obtain
\begin{align}
  \int_\omega^\Lambda \frac{d\Omega}{\pi}\int_\Sigma \frac{d^dk}{(2\pi)^d} \frac{\text{Im} \, G^R(\Omega,\vec k)}{\Omega} \leq \frac{1}{\omega n_B(\omega)} \frac{\langle n_0^2 \rangle_T}{\nu^2} \,. \label{eq:b1}
\end{align}
Here $n_B(\omega)$ is the Bose-Einstein distribution and $\langle \cdot \rangle_T$ is the expectation value in the thermal state.
For spinless fermions $n_0^2 = n_0 = c^\dagger_0 c_0$. For spin-half fermions $\langle n_0^2 \rangle_T$ only differs from $\langle n_0 \rangle_T$ in states with significant double occupancy (and, even then, only by an order one factor).

The second bound is stronger than (\ref{eq:b1}) at high frequencies, but weaker at low frequencies. It requires that the microscopic Hamiltonian be local and that the local terms in the Hamiltonian be bounded by a characteristic energy scale $\vep$, defined precisely in 
\S\ref{sec:methods}. This bound follows a similar logic to that in \cite{PhysRevLett.115.256803}, see also \cite{PhysRevX.9.041017}, but we have found it necessary to integrate over frequencies and momenta. The bound is
\be
\int_\omega^\Lambda \frac{d\Omega}{\pi}\int_\Sigma \frac{d^dk}{(2\pi)^d} \frac{\text{Im} \, G^R(\Omega,\vec k)}{\Omega} \leq \frac{1}{\omega n_B(\omega)} \frac{||n_0||^2}{\nu^2} \times \left\{
\begin{array}{cc}
1 & \text{for} \;\; \frac{\omega}{\vep} \leq 1  \\
\frac{\omega}{\vep} e^{2 (1-\omega/\vep)} & \text{for} \;\;  \frac{\omega}{\vep} \geq 1
\end{array}
\right.
\,.\label{eq:bf}
\ee
Here $|| \cdot ||$ is the operator norm. For $\omega \leq \vep$ the bound (\ref{eq:bf}) is weaker than (\ref{eq:b1}) but, as anticipated above, at large $\omega$ it is exponentially stronger. The large $\omega$ bound in (\ref{eq:bf}) has a stronger exponential falloff, by a factor of $e$, than the one given in \cite{PhysRevLett.115.256803}. Indeed, (\ref{eq:bf}) is the optimal exponential bound at large $\omega$.

\section{Spectral weight in the Drude-Kadanoff-Martin model}
\label{sec:drude}

We now choose an effective description of the charge dynamics at low energies and long wavelengths and ask how the microscopic bounds (\ref{eq:b1}) and (\ref{eq:bf}) constrain the parameters in the effective theory. A given microscopic model may or may not lead to the chosen effective theory. As we emphasised in the introduction, the logic is to determine the constraints that must be obeyed {\it within} a particular effective description.

A minimal effective description of charge dynamics at low energies and long wavelengths was given by Kadanoff and Martin \cite{KADANOFF1963419}
\be\label{eq:kad}
\frac{\text{Im} \, G^R(\Omega,\vec k)}{\Omega} = \frac{\chi D k^2}{\Omega^2 + (D k^2 - \tau \Omega^2)^2} \,.
\ee
Here $G^R(\Omega,\vec k)$ is the retarded Green's function for the charge density $n$, now understood in the continuum. The parameters in the model are the charge compressibility $\chi$, charge diffusivity $D$ and current relaxation time $\tau$. The charge compressibility is a thermodynamic quantity, given by the derivative of the charge density with respect to the chemical potential, $\chi \equiv \left.\pa n/\pa \mu\right|_T$. The optical conductivity is, using charge conservation $\nabla \cdot j + \dot n = 0$,
\be\label{eq:drude}
\sigma(\Omega) = \frac{\text{Im} \, G^R_{jj}(\Omega,0)}{\Omega} = \lim_{k \to 0} \frac{\Omega \, \text{Im} \, G^R(\Omega,\vec k)}{k^2} = \frac{\chi D}{1 + \tau^2 \Omega^2} \,.
\ee
This is the standard Drude form and for this reason we will refer to (\ref{eq:kad}) as the Drude-Kadanoff-Martin (DKM) model.\footnote{In the context of heat transport, this model is also known as the Maxwell–Cattaneo model.} One can understand (\ref{eq:kad}) as the minimal model that consistently combines Drude current relaxation (\ref{eq:drude}) with charge diffusion ($\Omega \approx - i D k^2$) at the lowest frequencies.
The dc conductivity in (\ref{eq:drude}) is given by the Einstein relation $\sigma_\text{dc} = \chi D$.

For an overview of the applicability of the Drude conductivity (\ref{eq:drude}) to real-world metals, see \cite{armitage}. The combined frequency and momentum dependence of (\ref{eq:kad}) exhibits a crossover from diffusion to linear dispersion as a function of increasing $k$. This crossover has been observed in a cold atomic realisation of a Hubbard-like system \cite{brown}. It is also instructive to note that (\ref{eq:kad}) can be derived from hydrodynamics in the limit of weakly broken translation invariance \cite{Davison:2014lua}. This limit highlights a key assumption in (\ref{eq:kad}), namely that there is a range of frequencies over which relaxation is described by the single timescale $\tau$. We discuss more general situations in \S\ref{sec:discussion}.

Whenever (\ref{eq:kad}) is the correct effective low energy description of charge transport, it will hold up to some cutoff in both frequency and momentum: $\omega \leq \Lambda$ and $k \leq \Lambda'$.
We are going to consider cases in which there is a well-defined Drude peak within the effective description (\ref{eq:drude}) and in which diffusion onsets within the effective description. That is, we make the further assumptions that $\frac{1}{\tau} \leq \Lambda$ and $\frac{1}{\sqrt{D \tau}} \leq \Lambda'$. These latter assumptions are not required for a bootstrap approach, but ensure that the effective theory has some teeth --- otherwise the spectral weight can be series expanded at small $\Omega$ or $k$ over the entire domain of validity.
On dimensional grounds $\sqrt{D \tau}$ defines a collective lengthscale. To emphasise this fact we introduce the notation
\be\label{eq:ell}
\ell \equiv \sqrt{D \tau} \,.
\ee
In weakly interacting systems $\ell$ is the mean free path.
More generally, it is the dynamical lengthscale defined within the DKM model (\ref{eq:kad}), independently of the existence of quasiparticles. 

We will use the DKM form (\ref{eq:kad}) in the bounds (\ref{eq:b1}) and (\ref{eq:bf}). The resulting spectral weight integrals, appearing on the left hand side of the bounds in \S\ref{sec:bounds}, are given in the methods section \ref{sec:methods}. In the methods we furthermore obtain a useful analytic bound on these integrals. The DKM spectral weight (\ref{eq:kad}) defines a classical effective field theory. Fluctuation (loop) effects in diffusive theories have been studied in recent papers \cite{Chen-Lin:2018kfl, Delacretaz:2020nit, Michailidis:2023mkd, Delacretaz:2023pxm}. See also \cite{PhysRevB.73.035113}. The fluctuations introduce non-analytic frequency dependence and can become important at low frequencies in low dimensions. Our bounds do not make any assumptions about analyticity of the Green's function at low frequencies. We will be working away from the lowest frequencies and will therefore assume that fluctuation effects are numerically small and can be neglected.

\section{Drude-Kadanoff-Martin model and microscopic energy scales}
\label{sec:micro}

\begin{figure}[h]
\centering
\includegraphics[width=0.7\textwidth]{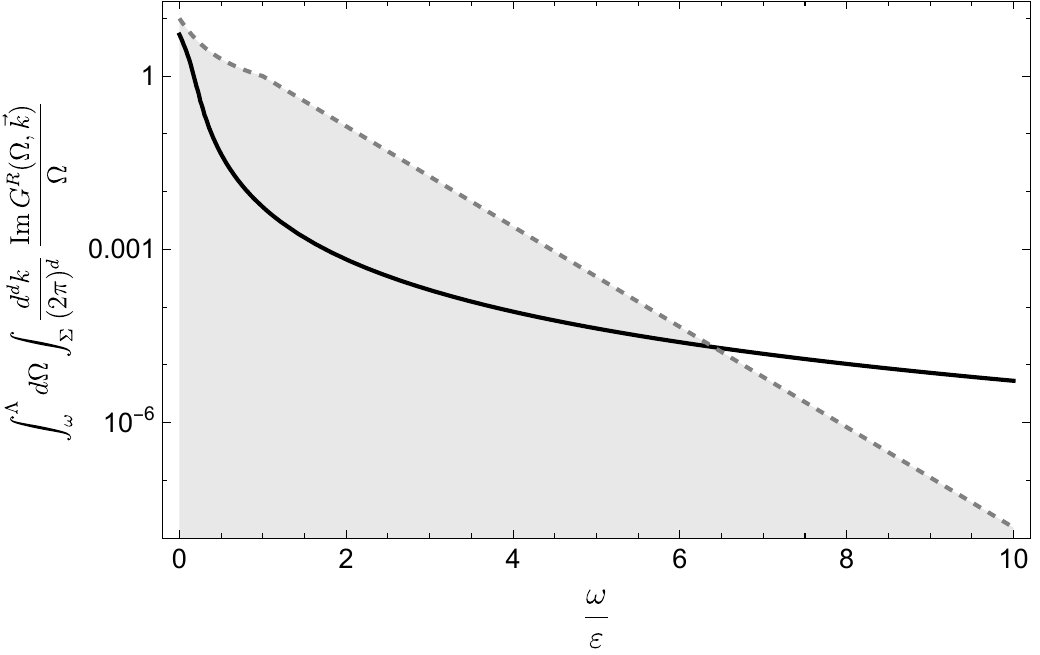}
\caption{The integrated spectral weight (solid curve) must lie within the shaded region, allowed by the upper bound (\ref{eq:bf}). The bound decays exponentially at large frequencies while the Drude spectral weight has a power-law decay. Thus the DKM form cannot extend to frequencies much greater than $\vep$. The plot uses the illustrative values $\frac{1}{\vep \tau} = \frac{1}{5}$, $\frac{T}{\vep} = \frac{1}{10}$ and $\frac{\Lambda}{\vep} = 50$ and takes $d=3$.}
\label{fig:expo}
\end{figure}

The effective DKM description is not expected to persist up to very high frequency scales. We can prove this `obvious' fact using the upper bound (\ref{eq:bf}). The exponential decay of (\ref{eq:bf}) at large frequencies entails that there cannot be any significant spectral weight above the microscopic energy scale $\vep$. In contrast, the integrated DKM spectral weight has a power law decay at large frequencies. Therefore, the cutoff on the DKM model must occur below the microscopic energy scale, $\Lambda \lesssim \vep$. This fact is illustrated in Fig.~\ref{fig:expo}, and we will give a precise statement shortly.
Before that, we make a further important comment about the Drude peak and microscopic scales. Suppose, as we have just argued is the case, that $\Lambda \lesssim \vep$. To obtain the bounds (\ref{eq:b1}) and (\ref{eq:bf}) we used positivity of
$\text{Im}\,G^R(\Omega,k)$ to `throw away' the spectral weight at energy scales above $\Lambda$. However, interband transitions can cause there to be a significant spectral weight at microscopic energy scales of order $\vep$. This microscopic spectral weight means that the effective low energy theory will be far from saturating the bounds. Similar issues arise in evaluating sum rules for multi-band systems \cite{armitage, RevModPhys.83.471}. For the purposes of the present work we will have in mind that the microscopic lattice model directly describes a single band system. 

To obtain a sharp statement relating $\Lambda$ and $\vep$ we evaluate the bound (\ref{eq:bf}) at $\omega = \frac{\Lambda}{2}$. For clarity we will make some simplifying assumptions about the parameter regime, none of which are essential: temperatures are well below the cutoff, so that the Bose-Einstein factor $n_B(\frac{\Lambda}{2}) \approx 1$, fermions are spin-half with $||n_0|| = 2$, and the relaxation rate $\frac{1}{\tau} \ll \Lambda$, well below the cutoff scale. Using the evaluation of the DKM spectral weight in the methods \S\ref{sec:methods}, the bound (\ref{eq:bf}) implies that
\be\label{eq:exp}
\frac{\gamma_d}{(\vep \tau)^3} \frac{\chi \nu^2 \vep}{\ell^d} \leq \left(\frac{\Lambda}{\vep}\right)^3 e^{-\Lambda/\vep} \,,
\ee
where the numerical prefactor $\gamma_d = \frac{7 e^{-2}}{12 \pi(d+2)}\frac{V_{d-1}}{(2\pi)^d}$.  Here $V_{d-1}$ is the volume of $S^{d-1}$. These prefactors are quite small.

To get a sense of (\ref{eq:exp}), and for later purposes, let us introduce the lattice spacing defined as
\be\label{eq:a}
a \equiv \nu^{1/d} \,.
\ee
We give a typical expression for $\chi$ in a metal in (\ref{eq:free}) below. For illustrative purposes, consider a metal with a large Fermi surface, so that in (\ref{eq:free}) we have $k_F \sim 1/a$ and $E_F \sim \vep$. The numerator on the left hand side of (\ref{eq:exp}) is then given by the lattice lengthscale $\chi \nu^2 \vep \sim a^d$. The bound (\ref{eq:exp}) therefore says that in order for the Drude peak to extend to frequencies well above the microscopic energy scale, i.e.~to have $\vep \ll \Lambda $, the mean free path $\ell$ must be exponentially longer than the microscopic lengthscale $a$ and/or the Drude lifetime $\tau$ must be exponentially longer than the microscopic timescale $1/\vep$. This is not commonly the case (although see e.g.~\cite{PhysRevLett.109.116401}).

\section{Bounds on the collective mean free path}
\label{sec:free}

\begin{figure}[h]
\centering
\includegraphics[width=0.7\textwidth]{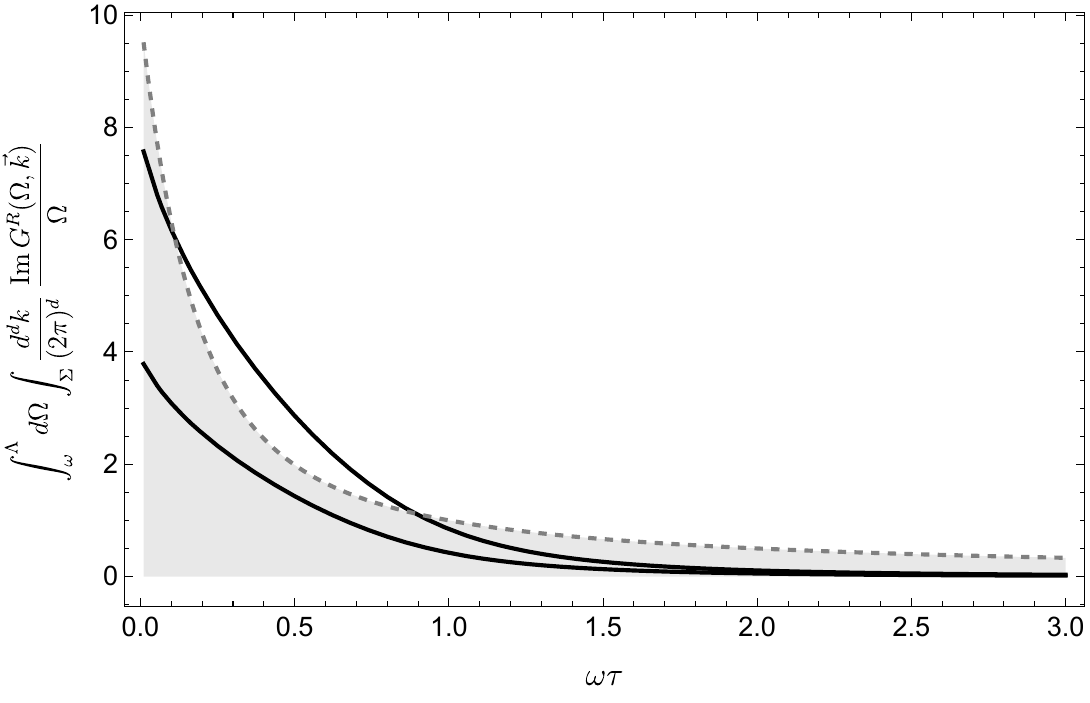}
\caption{The two solid curves are the integrated DKM spectral weight for different values of $\frac{\ell}{a}$. The lower solid curve obeys the inequality (\ref{eq:theb}) and the upper curve does not. Correspondingly, at intermediate frequencies $\omega \tau \lesssim 1$ the upper curve leaves the shaded region allowed by the bound (\ref{eq:b1}). In this plot we have worked in $d=2$, with $T\tau = \frac{1}{10}$ and $\Lambda \tau = 10$.}
\label{fig:planck}
\end{figure}

In the remainder, with the results of \S\ref{sec:micro} in mind, we will take the DKM model to extend up to $\Lambda \lesssim \vep$. This means that the exponential factor $e^{-2\omega/\vep} \approx 1$ in the bound (\ref{eq:bf}). That bound is then not stronger than the state-dependent bound (\ref{eq:b1}), so we will use (\ref{eq:b1}) in the following. The main result of this section will be a bound on the collective mean free path $\ell$, defined in (\ref{eq:ell}), in terms of the microscopic lengthscale $a$, defined in (\ref{eq:a}). The emergence of a bound on $\ell/a$ is illustrated in Fig.~\ref{fig:planck}.

We can obtain an explicit, analytic bound in cases where there is a well-defined peak within the effective DKM theory, i.e.~when $\frac{1}{\tau} \ll \Lambda$. This assumption simplifies the integrals involved, but relaxing this limit does not make a significant quantitative difference. In the methods \S\ref{sec:methods} we obtain from (\ref{eq:b1}) that
\be\label{eq:theb}
\frac{\beta_d}{1 - e^{-1/(T \tau)}}\leq \left(\frac{\ell}{a}\right)^d \frac{\tau}{\chi a^d} \langle n_0^2 \rangle_T \,,
\ee
where $\beta_d \equiv \frac{1}{\pi} \frac{V_{d-1}}{(2\pi)^d}\frac{1 - \frac{\pi}{4}}{d+2}$ is a numerical prefactor that can likely be improved. The bound (\ref{eq:theb}) is precise and general. To unpack the physical meaning of (\ref{eq:theb}) in the following few paragraphs we will make some estimates of the various quantities that appear.

The right hand side of (\ref{eq:theb}) is written as a product of dimensionless ratios. The first of these is the collective mean free path in terms of the lattice spacing, $\ell/a$. The second ratio is the current relaxation time divided by an inverse energy scale. To make this latter ratio more explicit, note that for non-interacting degenerate spinless fermions one has
\be\label{eq:free}
\chi = \frac{V_{d-1}}{(2\pi)^d} \frac{k_F^d}{2 E_F} \,, \qquad \frac{\langle n_0^2 \rangle_T}{\nu} = \frac{V_{d-1}}{(2 \pi)^d} \frac{k_F^d}{d} \qquad \text{(free fermions)} \,.
\ee
Here $k_F$ is the Fermi momentum and $E_F$ the Fermi energy. As $\chi$ and $\langle n_0^2 \rangle_T$ are thermodynamic quantities, the non-interacting expressions can be taken as a reasonable ballpark estimate for general systems. For example, in a Fermi liquid $\chi = \chi_\text{free}/(1 + F_0^S)$ is only modified numerically by the Landau parameter $F_0^S$.
With the estimates in (\ref{eq:free}) we can write the bound (\ref{eq:theb}) as
\be\label{eq:theb2}
\frac{d}{2}\frac{\beta_d}{1 - e^{-1/(T \tau)}}\leq \left(\frac{\ell}{a}\right)^d \tau E_F \,,
\ee
Typically we can expect that $\tau \propto \ell$, and hence (\ref{eq:theb2}) is a lower bound on the mean free path.

To make the bound (\ref{eq:theb2}) more transparent we will now consider five different regimes. 
\begin{figure}[h]
\centering
\includegraphics[width=\textwidth]{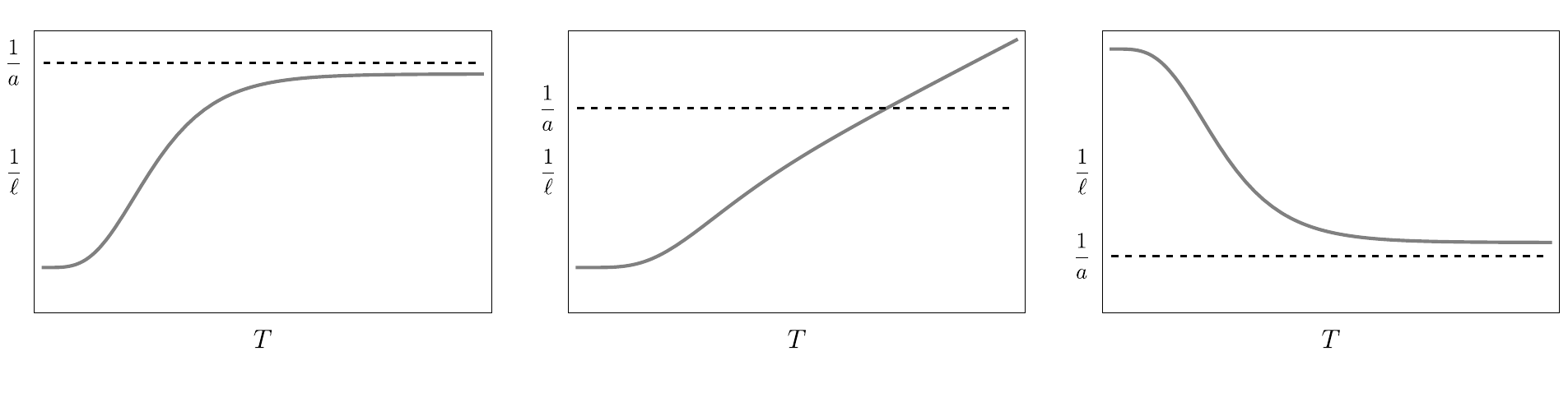}
\caption{Schematic evolution of the collective mean free path $\ell$ with temperature $T$, compared to the lattice spacing $a$. The inverse $1/\ell$ is a proxy for the resistivity. {\bf Left plot:} low temperature metal, intermediate temperature conventional or Planckian metal, high temperature saturation. The bound (\ref{eq:theb2}) is obeyed throughout. {\bf Middle plot:} Planckian bad metal at high temperatures, violating (\ref{eq:theb2}) in the regime where $\ell \lesssim a$. {\bf Right plot:} Insulator with $\ell < a$ at low temperatures, also violating (\ref{eq:theb2}).}
\label{fig:three}
\end{figure}
These regimes are illustrated schematically in Fig.~\ref{fig:three} and discussed below:

\paragraph{{\it (i)}  Low temperature metals and insulators:} Elastic disorder scattering is often dominant in metals at low temperatures. In this regime $\tau$ remains finite while $T \to 0$, leading to the `super-Planckian' limit $T \tau \ll 1$. Electrons undergoing predominantly disorder scattering are weakly interacting amongst themselves and hence $\ell$ and $\tau$ can be thought of as a single-particle mean free path and lifetime, related by the Fermi velocity $\ell = v_F \tau$. Thus we can
further estimate $\tau E_F \approx 2 \ell k_F \leq 2 \pi \ell/a$, leading to
\be\label{eq:supP}
\left(\frac{d \beta_d}{4 \pi}\right)^{\frac{1}{d+1}} \leq \frac{\ell}{a}\,. \qquad \text{(super-Planckian $T \tau \ll 1$)}
\ee
This inequality gives a version of the Mott-Ioffe-Regel bound on the mean free path \cite{RevModPhys.75.1085, Hussey}. Conversely, we may conclude that a system with a short low-temperature mean free path, violating (\ref{eq:supP}), cannot have a well-defined Drude peak in the sense we have described. This is consistent with the fact that such systems are insulators rather than metals \cite{RevModPhys.75.1085}, see the rightmost plot in Fig.~\ref{fig:three} (as opposed to the metallic leftmost plot).

\paragraph{{\it (ii)} Fermi liquids:} Inelastic scattering in a Fermi liquid at temperatures $T \ll E_F$ leads to the sub-Planckian current relaxation rate $1/\tau \sim T\times \frac{T}{E_F} \ll T$. In sub-Planckian regimes we obtain from (\ref{eq:theb2}) that
\be\label{eq:subP}
\frac{d \beta_d}{2} \frac{T}{E_F} \leq \left(\frac{\ell}{a}\right)^d \,. \qquad \text{(sub-Planckian $1 \ll T \tau$)}
\ee
With sub-Planckian scattering we may again think of $\ell$ and $\tau$ as single-particle properties related by the Fermi velocity so that, with a large Fermi surface, 
$\ell/a \sim \tau E_F/(k_F a) \sim \tau E_F \gg E_F/T$. Thus $a \ll \ell$ and hence (\ref{eq:subP}) will be obeyed. In particular, Fermi liquids are consistent with conventional Drude peaks. This is the low-intermediate temperature regime in the leftmost and middle plots of Fig.~\ref{fig:three}. We will consider high temperatures, above $E_F$, below. Finally, recall that conventional metals at intermediate temperatures are dominated by electron-phonon scattering rather than electron-electron scattering, but phonons are beyond our discussion in this paper.

\paragraph{{\it (iii)} Planckian scattering:} Formulae derived for weakly interacting quasiparticles are questionable in Planckian regimes, with $\tau \sim 1/T$ \cite{Hartnoll:2021ydi}. If we again estimate $\ell/a \sim \tau E_F/(k_F a) \sim \tau E_F \sim E_F/T$ in (\ref{eq:theb2}) then we find that the bound is satisfied for all $T \lesssim E_F$. Thus Planckian scattering is consistent with a conventional Drude peak up to $T \sim E_F$. Over these temperatures $a \lesssim \ell$, according to the estimates we have just made, and so the MIR bound is also obeyed.

\paragraph{{\it (iv)} Mean free path saturation at high temperatures:}
At high temperatures, $E_F \lesssim T$, the compressibility becomes temperature-dependent, $\chi \sim k_F^d/T$. This requires us to go back to the bound (\ref{eq:theb}). Here we consider sub-Planckian scattering. In this case the bound is now seen to require $a \lesssim \ell$. In particular, a conventional Drude peak is consistent with saturation of the mean free path at high temperatures to some value $\ell_\text{sat}$ that is greater than the lattice spacing. Saturation is illustrated in the leftmost plot of Fig.~\ref{fig:three}. Because of the temperature-dependent compressibility, the resistivity can continue to increase even while the mean free path saturates. Precisely this behaviour has been observed in a Hubbard-like system studied at very high temperatures by cold atom experiment \cite{brown} and by quantum Monte Carlo simulation \cite{mc}, and in a solvable model for this regime given in \cite{PhysRevLett.122.186601}. Indeed, as we noted above, dynamics consistent with the DKM form (\ref{eq:kad}) were reported in this regime \cite{brown}. On the other hand, the optical conductivity was found in \cite{PhysRevLett.122.186601} to exhibit well-defined, but unconventional peaks that were Gaussian rather than Lorentzian. These peaks indicate a departure, in the solvable model at least, from the frequency-dependence of the Drude form (\ref{eq:drude}). See \cite{PhysRevB.73.035113} for a broader discussion of ultra high-temperature optical response.

\paragraph{{\it (v)} High temperature bad metals:} In `bad metals' \cite{PhysRevLett.74.3253} the  mean free path does not saturate at high temperatures but continues to decrease below the lattice spacing $a$. High temperature bad metals are typically Planckian \cite{Bruin:2013qlp, Hartnoll:2014lpa, Hartnoll:2021ydi}, illustrated in the middle plot of Fig.~\ref{fig:three}. Using the high temperature behaviour $\chi \sim k_F^d/T$ together with Planckian dissipation $\tau \sim 1/T$ in the bound (\ref{eq:theb}) leads to $a \lesssim \ell$. It follows that Planckian bad metals are incompatible with conventional Drude peaks.
This conclusion is consistent with the usual understanding of bad metals as non-quasiparticle systems that necessarily transfer spectral weight out of the Drude peak, see e.g.~\cite{RevModPhys.75.1085, Hussey, 10.21468/SciPostPhys.3.3.025} for discussions and references.
On the other hand, a short mean free path extends the regime of validity of diffusion up to microscopic lengthscales \cite{Glorioso:2022poi}. The breakdown of the DKM form (\ref{eq:kad}) required by a violation of (\ref{eq:theb}) is thus expected to involve the nature of current relaxation rather than charge diffusion.\\

In summary, we have shown that systems with $\ell$ shorter than the lattice spacing
are not consistent with the DKM effective low energy description of transport. In particular, they cannot have conventional Drude peaks. This pertains to both low temperature disordered systems with very short mean free paths, which are in practice insulators, and to high temperature Planckian bad metals. On the other hand, `good metals' including Fermi liquids and metals with saturating mean free paths at high temperature are compatible with conventional Drude peaks (at least according to the bounds we have established here). We should emphasise that $\ell$ here is a collective mean free path that is well-defined within the effective low energy DKM model (\ref{eq:kad}). The version of the MIR bound that we have established is therefore sharply posed within that model.

\section{Discussion}
\label{sec:discussion}

In this work we have shown that upper bounds on the retarded Green's function for the charge density allow a bootstrap approach to establishing constraints on transport. We will end with a discussion of some of the many future directions that this approach should enable.

\paragraph{Beyond the DKM model.} The DKM model is the ‘harmonic oscillator’ of charge dynamics. As we have noted in the introduction, many interesting metals do not exhibit textbook Drude peaks. Their electrodynamics requires modelling instead with, for example, the so-called extended Drude model \cite{armitage, RevModPhys.83.471}. The results in \S\ref{sec:free} rely on the specific single-relaxation-time structure of the DKM model in (\ref{eq:kad}). In Fig.~\ref{fig:planck} we can see that redistribution of spectral weight from intermediate to higher frequencies, resulting in a peak that falls off more slowly than a conventional Drude, helps to satisfy the bound. That said, our bounds (\ref{eq:b1}) and (\ref{eq:bf}) are independent of this model and must be obeyed by any more complicated form of charge response. In the following paragraphs we discuss further model-independent constraints. In particular, the constraints (\ref{eq:bn}) and (\ref{eq:LR_analyticity_momentum_space}) below pertain beyond the DKM model.

\paragraph{Analyticity in frequency.} As we noted in the introduction, analyticity plays a key role in the bootstrap of scattering amplitudes. The corresponding statement in our context is that the retarded Green's function is analytic in the upper half complex frequency plane. A well-known consequence of this analyticity are the Kramers-Kronig relations between the real part of the Green's function on the real frequency axis and the integral of the imaginary part. The Kramers-Kronig relations are a useful starting point to prove sum rules, but are not so useful for establishing bounds because the integral involves a principal value that is not easily bounded. In fact, there is no general bound on the real part of the Green's function on the real axis. This is a departure from the analogy with scattering amplitudes, in which $|1 + i T_l(s)| \leq 1$ implies that the real and imaginary parts of $T_l(s)$ are both bounded.

A more fruitful approach is to consider the Green's function on the positive imaginary frequency axis. Imaginary frequencies are equivalent to smearing over real frequencies, as a simple contour integral in the upper half plane gives, with $\a > 0$,
\be\label{eq:kk2} 
G^R(i \a,k) = \frac{2}{\pi} \int_0^\infty \frac{\Omega \,\text{Im} \, G^R(\Omega,k)}{\Omega{}^2 + \a^2} d\Omega \,.
\ee
Here we used $\text{Im} \, G^R(\Omega,k) = - \text{Im} \, G^R(-\Omega,k)$, using time-reversal symmetry to restore $k \to -k$ after complex conjugation \cite{Hartnoll:2009sz}, as well as the fact that the charge density Green's function vanishes at large $\Omega$ so that there is no contribution from the contour at infinity. While imaginary frequencies are not directly physical, if $\alpha \leq \Lambda$ then we can evaluate $G^R(i \a,k)$ using the effective low energy theory. Following essentially the same steps as led to (\ref{eq:b1}) we can bound
\begin{align}\label{eq:bn}
g^R(i \a) \leq H\left(\frac{\a}{T}\right) \, \frac{\left\langle n_0^2 \right\rangle_T}{\a \, \nu^2} \,.
\end{align}
Here we introduced the momentum-integrated Green's function and the dimensionless function
\be
g^R(i \a) \equiv \int_\Sigma \frac{d^dk}{(2\pi)^d} G^R(i \a,k) \,, \qquad H(x) \equiv \max_{0 \leq y} \frac{2 y}{1 + y^2} (1 - e^{- x y}) \,.
\ee
An advantage of (\ref{eq:bn}) relative to our previous bounds is that it does not involve an integral over frequencies. This will be helpful in cases where the effective low energy Green's function is more complicated than the DKM form. For the DKM model
\be
G^R(i \a,k) = \frac{\chi D k^2}{\a + \tau \a^2 + D k^2} \,,
\ee
and one obtains similar constraints to those we found in \S\ref{sec:free}.

In \eqref{eq:kk2} we used the simple function $\frac{\Omega}{\Omega^2 + \alpha^2}$, which is odd under $\Omega \rightarrow -\Omega$ and has a pole at $\Omega = i \alpha$. The bound (\ref{eq:bn}) can be improved by optimising over a suitable class of functions with these properties. This is the idea behind the dual S-matrix bootstrap \cite{Cordova:2019lot, Guerrieri:2020kcs}.

\paragraph{Causality and analyticity in momentum space.}
The exponential lightcone enforced by the Lieb-Robinson bound \cite{Lieb:1972wy, Chen_2023} leads to a domain of analyticity in momentum space also. Let us sketch how this works. The Lieb-Robinson bound states that the commutator of two local operators decays exponentially outside the light-cone 
\be
\label{eq:LR_bound}
|| [ A(\vec x,t), A(0,0) ]|| \leq C e^{-\kappa(x-v t)} \,.
\ee
Here $v$ is the Lieb-Robinson velocity and $\kappa$ is a microscopic inverse lengthscale. The decay of the operator norm in (\ref{eq:LR_bound}) bounds the retarded Green's function, a fortiori. The Fourier transform of the Green's function must then be analytic in momentum space for complex $\omega$ and $\vec k$ such that
\be
\label{eq:LR_analyticity_momentum_space}
\text{Im} (\omega) > v \, | \text{Im} (\vec k) | \quad \text{and} \quad | \text{Im} (\vec k) | < \kappa \,,
\ee
and real parts arbitrary. A quick proof follows from considering the Fourier transform of (\ref{eq:LR_bound}) at complex $\omega$ and $\vec k$. The Fourier transformed Green's function is assumed to exist for real $\omega$ and $k$, at least in the sense of tempered distributions, and the question is whether the imaginary parts of these quantities improve the convergence of the integral.  For the retarded Green's function we have $t>0$. Outside of the future lightcone, i.e.~for $0 < v t < x \equiv |\vec x|$, we have
\be\label{eq:trans}
   e^{i \omega t - i \vec k. \vec x} e^{-\kappa(x-v t)} \leq e^{- \left[\text{Im}(\w) - v \k \right] t} e^{\left[|\text{Im} (\vec k)| -\kappa \right] x} e^{i \text{Re}(\w) t - i   \text{Re} (\vec k) \cdot \vec x}  \,.
\ee
Here we used $\text{Im} (\vec k) \cdot \vec x \leq |\text{Im} (\vec k)| x$. From (\ref{eq:trans}) we see that all directions outside of the lightcone are damped at large $t$ and $x$ if and only if the two conditions in (\ref{eq:LR_analyticity_momentum_space}) hold, as these give damping in the limiting cases of $t = x/v$ and $t=0$. Inside the future lightcone the bound (\ref{eq:LR_bound}) is very weak and the commutator norm is better bounded by the constant $2 ||A||^2$. Using $v t > x$ we have
\be\label{eq:in}
   e^{i \omega t - i \vec k. \vec x} \leq e^{- \left[\text{Im}(\w) - v |\text{Im} (\vec k)| \right] t} e^{i \text{Re}(\w) t - i   \text{Re} (\vec k) \cdot \vec x}  \,.
\ee
The first condition in (\ref{eq:LR_analyticity_momentum_space}) is enough to ensure that these terms are damped as $t \to \infty$.

In \cite{Heller:2022ejw, Heller:2023jtd} the first of the inequalities in (\ref{eq:LR_analyticity_momentum_space}), following from a strict relativistic lightcone, was used to rigorously derive the diffusivity upper bound of \cite{Hartman:2017hhp}. The second inequality in (\ref{eq:LR_analyticity_momentum_space}) extends that derivation to the case of an exponential lightcone, now with the additional assumption that $1/\tau < v \kappa$ so that causality is enforced over the timescale at which local equilibration occurs. It is interesting that we can read our bound (\ref{eq:theb}) as a lower bound on diffusivity, re-instating $\ell = \sqrt{D \tau}$. Combing that bound with the causality upper bound, and only keeping track of factors of $\tau$ and $D$, gives $\tau^{-1-2/d} \lesssim D \lesssim \tau$. Taken together these inequalities imply a lower bound on $\tau$ itself. It seems clear, therefore, that incorporating causality into the upper growth bound in \S\ref{sec:bounds} will lead to new constraints on low energy dynamics.

\paragraph{Continuum UV completions.} There has been significant interest in constraining the low energy and long wavelength dynamics of conserved quantities in theories with a continuum short distance completion, such as a bound on the viscosity of the quark-gluon plasma \cite{Kovtun:2004de}. Both of our bounds (\ref{eq:b1}) and (\ref{eq:bf}) used the discreteness of the lattice completion in an essential way. This is seen, for example, in the factors of $\nu$ appearing in the bounds. A continuum version of (\ref{eq:b1}) may well be possible, but will have to make sense of the composite operator $n(x)^2$. It is interesting that relativistic hydrodynamics is typically in need of a Drude-like relaxation time $\tau$ in order to render the theory causal (see e.g.~\cite{Romatschke:2009im}). The effective long wavelength description of the quark-gluon plasma, then, shares some aspects of the DKM model that we have discussed. Furthermore, while the exponential decay in our bound (\ref{eq:bf}) is directly tied to bounded lattice interactions, it is remarkable that \cite{Romatschke:2009ng} found a similar exponential falloff in the {\it temperature dependent} part of the spectral density in $\mathcal{N}=4$ SYM at strong coupling. See Fig.~2 therein. Potentially, then, it may be possible to establish a continuum version of (\ref{eq:bf}).

\paragraph{Solving strongly correlated transport.} A more ambitious bootstrap goal is to `solve' a specific model by strongly bounding observables on both sides. We noted above that incorporating causality constraints will likely lead to both upper and lower bounds. Another route to additional bounds are constraints from fluctuations, the most sophisticated discussion of fluctuation bounds to date is \cite{Delacretaz:2023pxm}. Fluctuation corrections to transport are typically inversely proportional to the classical transport coefficients. Intuitively, this is because transport coefficients are proportional to the mean free path $\ell$ whereas fluctuations are controlled by the local thermalisation volume $1/\ell^d$. A bound on the sum of classical and fluctuation contributions is then a nonlinear constraint on $\ell$ that may lead to both upper and lower bounds.

Because lattice models are not in themselves highly constrained, unlike e.g.~CFTs \cite{Poland2016}, it will also be necessary to have a systematic procedure to improve bounds by incorporating more information from the microscopic theory. One approach is to package this information recursively into Lanczos coefficients, as in e.g.~\cite{PhysRevX.9.041017}. These coefficients can likely to be used to systematically strengthen the bounds in \S\ref{sec:bounds}, especially at low frequencies.

\section{Methods}
\label{sec:methods}

\subsection{First bound}

The spectral density for the Green's function (\ref{eq:gr}) is
\be\label{eq:spec}
\text{Im} \, G^R(\omega,\vec k) = \sum_{m,n} \frac{e^{- E_n/T}}{{\mathcal Z}_T} \frac{\pi \left|\langle n|n_0| m \rangle \right|^2}{n_B(\omega)} \frac{(2 \pi)^d}{\nu^2} \delta^{*}(\vec k + \vec p_n - \vec p_m) \delta\left(\omega + E_n - E_m \right) \,.
\ee
Here the Bose-Einstein factor $n_B(\omega) = 1/(1 - e^{- \omega/T})$, $|m\rangle$ is a simultaneous eigenstate of the Hamiltonian and lattice translations, obeying
\be
H |m\rangle = E_m |m\rangle \,, \qquad  T_{\vec x_\a} |m\rangle = e^{i \vec p_m \cdot \vec x_\a} |m\rangle \,,
\ee
and $\delta^*$ is a $d$-dimensional delta function supported on the dual (momentum-space) lattice. This delta function, and an additional factor of $1/\nu$, arise from Poisson summation. We can take the $\vec p_m$ and $\vec k$ to be in the Brillouin zone.

Integrating the spectral weight (\ref{eq:spec}) gets rid of the explicit delta functions. Firstly, we integrate the momentum over some sub-region $\Sigma$ of the Brillouin zone to obtain
\be\label{eq:two}
\int_\Sigma \frac{d^dk}{(2\pi)^d} \text{Im} \, G^R(\omega,\vec k) \leq \frac{\pi}{\nu^2}\sum_{m,n} \frac{e^{- E_n/T}}{{\mathcal Z}_T} \frac{\left|\langle n|n_0| m \rangle \right|^2}{n_B(\omega)} \delta\left(\omega + E_n - E_m \right) \,.
\ee
The inequality follows from the fact, manifest in (\ref{eq:spec}), that $0 \leq \text{Im} \, G^R(\omega,k)$ for all $k$ and $0\leq\omega$. The integral over $\Sigma$ is therefore less than or equal to the integral over the entire Brillouin zone. Integrating $k$ over the entire Brillouin zone picks out a single one of the momentum delta functions in (\ref{eq:spec}) for every pair $m,n$ in the sum. The reason we restrict the integral to $\Sigma$ is that we will only wish to input the form of the Green's function at suitably low momenta.

To deal with the delta functions remaining in (\ref{eq:two}) we further integrate over frequencies.
Again, we only want to use the Green's function at low frequencies, and hence impose a cutoff $\Lambda$ on the integral.
Consider, with $0 \leq \omega$,
\begin{align}
  \int_\omega^\Lambda \frac{d\Omega}{\pi}\int_\Sigma \frac{d^dk}{(2\pi)^d} \frac{\text{Im} \, G^R(\Omega,\vec k)}{\Omega} & \leq
  \frac{1}{\omega n_B(\omega)} \frac{1}{\nu^2} \sum_{\substack{m,n \\ \omega \leq E_m - E_n \leq \Lambda}}\frac{e^{- E_n/T}}{{\mathcal Z}_T} \left|\langle n|n_0| m \rangle \right|^2 \label{eq:b0}\\
  & \leq \frac{1}{\omega n_B(\omega)} \frac{1}{\nu^2} \sum_{m,n}\frac{e^{- E_n/T}}{{\mathcal Z}_T} \left|\langle n|n_0| m \rangle \right|^2 \\
  & = \frac{1}{\omega n_B(\omega)} \frac{\langle n_0^2 \rangle_T}{\nu^2} \,. 
\end{align}
In the first line we used the fact that $1/(\omega n_B(\omega))$ is a decreasing function of $\omega$, allowing us to take this factor outside of the integral. In the second line we added back in all of the $m,n$ pairs that do not obey the energy condition in the first line --- because all terms are positive this can only cause the expression to increase. In the final line we performed the sum over $m$ to obtain the answer in terms of the thermal expectation value of the charge density squared. 

\subsection{Second bound}

Recall firstly that
\be
\langle n|[H,n_0]| m \rangle = (E_n - E_m) \langle n|n_0| m \rangle \,.
\ee
Using this result $\k$ times in (\ref{eq:b0}), together with the bound on $E_m - E_n$ in the sum, we obtain
\begin{align}
\int_\omega^\Lambda \frac{d\Omega}{\pi}\int_\Sigma \frac{d^dk}{(2\pi)^d} \frac{\text{Im} \, G^R(\Omega,\vec k)}{\Omega} & \leq
  \frac{1}{\omega^{1+2\k} n_B(\omega)} \frac{1}{\nu^2} \sum_{\substack{m,n \\ \omega \leq E_m - E_n \leq \Lambda}}\frac{e^{- E_n/T}}{{\mathcal Z}_T} \Big|\langle n|\overbrace{\, [H,[H, \cdots \, , n_0]]}^{\k\ \text{commutators}} | m \rangle \Big|^2 \nonumber \\
  & \leq \frac{1}{\omega^{1+2\k} n_B(\omega)} \frac{1}{\nu^2} \sum_{n}\frac{e^{- E_n/T}}{{\mathcal Z}_T} \langle n| \Big|\overbrace{\, [H,[H, \cdots \, , n_0]]}^{\k\ \text{commutators}} \Big|^2 | n \rangle \\
  & \leq \frac{1}{\omega^{1+2\k} n_B(\omega)} \frac{1}{\nu^2} \Big|\Big|\overbrace{\, [H,[H, \cdots \, , n_0]]}^{\k\ \text{commutators}} \Big|\Big|^2 \,. \label{eq:c0}
\end{align}
Going from the first to the second line we have repeated the steps below (\ref{eq:b0}): the bound on $E_m - E_n$ has been relaxed, leading to a further inequality, and then the sum over $m$ has been performed. In the second line $|A|^2 \equiv A A^\dagger$. The third line follows from firstly using the definition of the operator norm to note that $\langle n|AA^\dagger|n\rangle \leq ||A||^2$ and then performing the sum over $n$. With $\k=0$ the final line (\ref{eq:c0}) is almost the same as (\ref{eq:b0}), but weaker because $\langle n_0^2\rangle_T \leq||n_0||^2$.
However, we will now use facts about operator growth to bound the commutator at large $\k$. This leads to an improved bound at large $\omega$.

Write the local Hamiltonian as $H = \sum_\alpha h_\alpha$, with $||h_\alpha|| \leq h$ and $h_\alpha$ supported within a ball of radius $R$. Suppose also that $n_0$ has support with a ball of the same radius. Then, it was shown in \cite{PhysRevLett.115.256803} that
\be\label{eq:lr}
\Big|\Big|\overbrace{\, [H,[H, \cdots \, , n_0]]}^{\k\ \text{commutators}} \Big|\Big| \leq \kappa! \vep^\kappa ||n_0||\qquad \text{with} \qquad \vep \equiv 2 h \frac{b_d (2R)^d}{\nu} \,.
\ee
Up to geometric factors, $\vep$ is a characteristic energy scale of the local terms in the Hamiltonian. In the definition of $\vep$ we have used $b_d$ to denote the volume of a unit $d$-dimensional ball and recall that $\nu$ is the unit cell volume.
The proof is straightforward: {\it (1)} Insert $H = \sum_\alpha h_\alpha$ into the commutators to obtain a sum of terms of the form $[h_{\a_\k},[h_{\a_{\k-1}}, \cdots , [h_{\a_1}, n_0]]]$, {\it (2)} the norm of any such term that is nonzero is bounded above by $(2 h)^\k ||n_0||$, with the factors of 2 coming from the commutators, {\it (3)} there are at most $\k! \left[b_d (2R)^d/\n\right]^\kappa$ nonzero terms --- as we now explain. For a term to be nonzero, the support of the $h_{\a_\k}$ operator must overlap with the support of one of the $\kappa$ operators $\{h_{\a_{\k-1}},\ldots, h_{\a_1}, n_0\}$. That is, the centre of the radius $R$ ball containing $h_{\a_\k}$ must be within a distance $2R$ of the centre of one of the $\k$ radius $R$ balls containing the other operators. There are at most $\k \, b_d (2 R)^d/\nu$ sites for which this is true. The same argument applies to the $h_{\a_{\k-1}}$ operator, which now has $\kappa - 1$ other operators that it can be close to. Iterating this procedure leads to the $\kappa!$ term in (\ref{eq:lr}).

We may now use (\ref{eq:lr}) in (\ref{eq:c0}) and make the choice $\kappa = \lfloor \frac{\omega}{\vep} \rfloor$. The floor function and the factorial can be removed with the following chain of inequalities
\be\label{eq:ineq}
\frac{\lfloor x \rfloor!}{x^{\lfloor x \rfloor}} \leq \frac{\Gamma(x+1)}{x^x} \leq x^{1/2} e^{1-x} \,, 
\ee
where the second inequality requires $x \geq 1$. Using (\ref{eq:ineq}) we see that 
(\ref{eq:c0}) implies the bound
\be
\int_\omega^\Lambda \frac{d\Omega}{\pi}\int_\Sigma \frac{d^dk}{(2\pi)^d} \frac{\text{Im} \, G^R(\Omega,\vec k)}{\Omega} \leq \frac{1}{\omega n_B(\omega)} \frac{||n_0||^2}{\nu^2} \times \left\{
\begin{array}{cc}
1 & \text{for} \;\; \frac{\omega}{\vep} \leq 1  \\
\frac{\omega}{\vep} e^{2 (1-\omega/\vep)} & \text{for} \;\;  \frac{\omega}{\vep} \geq 1
\end{array}
\right.
\,.
\ee

\subsection{Computation of the DKM spectral weight}

The DKM form holds up to the cutoffs $\Lambda$ and $\Lambda'$. However, it is more manageable and sufficient for our purposes to take the momentum integral to run only over the region $k \leq \frac{1}{\ell}$, rather than up to the cutoff $\Lambda'$. This restriction means that the numerical coefficients in our bootstrap bounds could likely be strengthened. The left hand side of the bounds in \S\ref{sec:bounds} then becomes
\be\label{eq:left}
 V_{d-1} \int_\omega^\Lambda \frac{d\Omega}{\pi} \int_0^{\frac{1}{\ell}} \frac{k^{d-1} dk}{(2\pi)^d} \frac{\text{Im} \, G^R(\Omega,k)}{\Omega} = \frac{1}{\pi} \frac{V_{d-1}}{(2\pi)^d} \frac{\chi}{\ell^d} F_d(\omega \tau,\Lambda \tau) \,,
\ee
where $V_{d-1}$ is the volume of $S^{d-1}$ and the dimensionless function
\be\label{eq:Fd}
F_d(z, \lambda) \equiv \int_{z}^\lambda dy \int_0^{1} dx \frac{x^{d+1}}{y^2 + (x^2 - y^2)^2} \,.
\ee
For later use we note here a lower bound on (\ref{eq:Fd}) that holds whenever $z \geq 1$. In this case we have $y \geq x$ everywhere in the domain of integration and hence the uniform bound $(x^2-y^2)^2 \leq y^4$. Using this bound in (\ref{eq:Fd}) renders the integrals elementary and we obtain
\be\label{eq:fbound}
\frac{1}{d+2}\left(\frac{1}{z} + \arctan{z} - \frac{1}{\lambda} - \arctan{\lambda}\right) \leq F_d(z,\lambda) \qquad (\text{for} \; z \geq 1)\,.
\ee
For $z \leq 1 \leq \lambda$ one obtains a lower bound by setting $z \to 1$ on the left hand side of (\ref{eq:fbound}). The lower bound in (\ref{eq:fbound}) accurately captures the large $z$ behaviour of the right hand side, but is weaker by numerical factors, which are $\lesssim 10$, at low values of $z$. We will use (\ref{eq:fbound}) as a manageable approximation to the integral (\ref{eq:Fd}) and again note that the numerical factors in our bounds below can likely be strengthened. In particular, for the discussion in \S\ref{sec:micro} we use that at 
large $\lambda$ one has $\frac{1}{d+2} \frac{7}{3 \lambda^3} \leq F_d(\frac{\lambda}{2},\lambda)$. 

To obtain (\ref{eq:theb}) we restricted to cases where $\frac{1}{\tau} \ll \Lambda$. This allows us to take $\lambda \to \infty$ in (\ref{eq:Fd}). To obtain the strongest constraint, we maximise the bound (\ref{eq:b1}) over all frequencies $\omega$:
\be\label{eq:max}
\frac{1}{\pi} \frac{V_{d-1}}{(2\pi)^d} \max_{0 \leq \omega} \Big[ \omega\tau  F_d(\omega \tau,\infty) n_B(\omega) \Big]\leq \left(\frac{\ell}{a}\right)^d \frac{\tau}{\chi a^d} \langle n_0^2 \rangle_T \,.
\ee
Here we used the definition (\ref{eq:a}) for $a$. The function being maximised in (\ref{eq:max}) goes to zero at large frequencies. That is to say, at large frequencies the integrated spectral weight always vanishes faster than the $1/\omega$ on the right hand side in the bound (\ref{eq:b1}). Any tension with the bound must therefore come at intermediate frequencies, as we illustrated in Fig.~\ref{fig:planck}.
We will use the lower bound (\ref{eq:fbound}) for $F_d$. With this expression, the maximum is reached at $\omega \tau = 1$ and below this frequency the maximand is constant. We thus obtain (\ref{eq:theb}).

\section*{Acknowledgements}

This work has been partially supported by STFC consolidated grant ST/T000694/1. SAH is partially supported by Simons Investigator award \#620869. SDC \& AH are supported by ``Exotic High Energy Phenomenology'' (X-HEP), a project funded by the European Union -- Grant Agreement n.~101039756 (PI: J.~E.~Mir\'o). Views and opinions expressed are however those of the author(s) only and do not necessarily reflect those of the European Union or the ERC Executive Agency (ERCEA). Neither the European Union nor the granting authority can be held responsible for them. SDC acknowledges Aspen Center for Physics, which is supported by a grant from the Simons Foundation (1161654, Troyer), the Galileo Galilei Institute for Theoretical Physics and the INFN for partial support during the completion of this work.

\providecommand{\href}[2]{#2}\begingroup\raggedright\endgroup

\end{document}